\documentclass[10pt]{iopart}
\usepackage{iopams}
\usepackage[utf8]{inputenc}
\usepackage{color}
\usepackage{overpic}
\usepackage{siunitx}
\usepackage{soul}
\usepackage{cite}

\newcommand{\DREAM}{\textsc{Dream}}
\usepackage{graphicx}
\usepackage[numbers]{natbib}
\newcommand{\eqref}[1]{(\ref{#1})}

\usepackage{float}
\usepackage{graphicx}
\graphicspath{{./figure/}}

\usepackage{pgfplots}
\pgfplotsset{compat=1.15}
\usepackage{pgfplotstable} 
\usepgfplotslibrary{external} 
\usepackage[breaklinks=true]{hyperref}
\hypersetup{
  unicode=false,            pdftoolbar=true,          pdfmenubar=true,          pdffitwindow=false,       pdfstartview={FitH},      pdfsubject={Subject},     pdfcreator={Creator},     pdfproducer={Producer},   pdfkeywords={keyword1} {key2} {key3},   pdfnewwindow=true,        colorlinks=true,          linkcolor=blue,           citecolor=blue,           filecolor=blue,           urlcolor=blue           }

\begin{document}
\title[O.~Vallhagen et al, Two-stage SPI in Tokamak Disruptions]{Effect of Two-Stage Shattered Pellet Injection on Tokamak Disruptions
  }
\author{O.~Vallhagen$^1$, I.~Pusztai$^1$, M.~Hoppe$^1$, S.~L.~Newton$^2$, T.~F\"ul\"op$^1$}
\address{$^1$ Department of Physics, Chalmers University of Technology,
  SE-41296 Gothenburg, Sweden}
\address{$^2$ Culham Centre for Fusion Energy, Abingdon, Oxon OX14 3DB, United Kingdom}
    \ead{vaoskar@chalmers.se}
\begin{abstract}
An effective disruption mitigation system in a tokamak reactor should limit the exposure of the wall to localized heat losses and to the impact of high current runaway electron beams, and avoid excessive forces on the structure. We evaluate  with respect to these aspects a two-stage deuterium-neon shattered pellet injection in an ITER-like plasma, using simulations with the \DREAM\ framework [M.~Hoppe et al (2021) Comp.~Phys.~Commun.~268, 108098]. To minimize the obtained runaway currents an optimal range of injected deuterium quantities is found. This range is sensitive to the opacity of the plasma to Lyman radiation, which affects the ionization degree of deuterium, and thus avalanche runaway generation. The two-stage injection scheme, where dilution cooling is produced by deuterium before a radiative thermal quench caused by neon, reduces both the hot-tail seed and the localized transported heat load on the wall. However, during nuclear operation, additional runaway seed sources from the activated wall and tritium make it difficult to reach tolerably low runaway currents.        
\end{abstract}
\ioptwocol
\section{Introduction}
The sudden loss of the energy confined in fusion plasmas
during off-normal events, called disruptions, presents one of the most severe
threats to the future of fusion energy based on the tokamak design. An
efficient disruption mitigation system is therefore of utmost
importance for future high-current devices such as ITER. The
potentially greatest threat to be mitigated is posed by large currents
carried by highly energetic \emph{runaway electrons}, whose number increases exponentially during the runaway avalanche \cite{RP97} from any small seed population, and
which 
may cause severe damage upon wall impact. The disruption mitigation
system must also ensure sufficiently homogeneous deposition of the
thermal energy on the plasma-facing components, and avoid excessive
forces on the machine resulting from currents flowing in the surrounding
structures.

The mitigation method currently envisaged is the injection of massive quantities of material
when an emerging disruption is detected, aiming to
better control the plasma cooling and energy dissipation.
Conventionally, the material is delivered as a gas puff 
from a pressurized vault \cite{ITERDisruptions}. This technique, while comparatively simple, has a number of disadvantages. The injected gas ionizes rapidly when exposed to the initially hot plasma, and so becomes tied to the magnetic field, substantially slowing the spread of the gas through the plasma. Moreover, the change in plasma profiles during the gas injection can accelerate the growth of plasma instabilities, allowing the disruption to progress before injected material has reached all parts of the plasma.

Another approach, that can provide better penetration, is to inject material in the form of a solid, cryogenic pellet. The exposure to the plasma causes the pellet to ablate and deposit its content along its trajectory. The ablation 
can be made more efficient by shattering the pellet into smaller shards before it enters the plasma. This shattered pellet injection (SPI)  technique has been chosen as the baseline for the disruption mitigation system at ITER \cite{LehnenIAEA}.

The details of the design and the operation parameters for the disruption mitigation system in reactor-scale devices remain an open question. 
A complexity arises as different aspects of the mitigation generate potentially conflicting requirements. For example, radiative dissipation of the thermal energy favours large injected quantities, which may result in an unacceptably fast quench time of the plasma current, due to the high plasma resistivity at low temperature. In addition, rapid cooling increases the hot-tail runaway seed \cite{Chiu_1998,Harvey2000}.

A method suggested recently to circumvent these issues is to divide the injection into two stages \cite{Nardon_2020}.
First, injection of a large amount of pure deuterium is used to cool the plasma through dilution, by a factor 10--100, without significantly perturbing the magnetic field configuration or radiating any substantial amount of thermal energy. A few milliseconds later a smaller neon injection follows, which radiatively dissipates the thermal energy. The division of the cooling into two steps gives the hot electrons (in the tail of the electron distribution) time to equilibrate at an intermediate temperature before the runaway generation is initiated, potentially suppressing the hot-tail runaway generation. In addition, if the plasma perturbation destroying the plasma confinement is initiated at a lower temperature, the slower thermal motion reduces the thermal energy transport. Instead, a larger fraction of the thermal energy can be dissipated through radiation, reducing the danger of localised wall hot spots. 

The results presented in reference~\cite{Nardon_2020} indicate that it is indeed possible to substantially cool the plasma through dilution by a deuterium shattered pellet injection, without destroying the plasma confinement. However the outcome and optimisation of this two-stage injection scheme, particularly the runaway generation, have not yet been studied in detail.

Recent simulations of plasma shutdown scenarios, focused on the dynamics during the current quench, indicate that material
injection can lead to high runaway currents in ITER
\cite{MS2017,ITER_OV}. While a massive increase in the electron density appeared as a promising route to suppress the runaway avalanche
\cite{MS2017}, the substantial
recombination in such scenarios results in a net enhancement of the final runaway
current \cite{ITER_OV}. These studies used a simplified instantaneous impurity deposition
with a flat radial profile and prescribed thermal quench dynamics.

 The hot-tail runaway seed generated during the thermal quench varies over many orders of
 magnitude, depending on the injected densities and seed losses arising from perturbations of the
 magnetic field \cite{AB17,Svenningsson2021}. Regions of
 parameter space with sufficiently small hot-tail seeds have been found previously, which are not expected
 to result in an excessive final runaway current despite the strong
 avalanche gain \cite{Svenningsson2021}. However, these studies did
 not model the consistent build-up of the injected density following SPI and
 the subsequent current quench dynamics.

In this work we investigate the effect of SPI on runaway generation in
a tokamak disruption. We use the recently developed numerical tool
\DREAM\ (Disruption Runaway Electron Analysis Model)
\cite{DREAMPaper}, which is capable of self-consistently calculating
the time evolution of the background plasma properties, the electron
momentum distribution and the runaway current during a mitigated
tokamak disruption. In particular, we simulate two-stage SPI in an ITER-like
setting, and assess how the disruption mitigation performance depends
on the amount of injected deuterium and neon. This assessment is
based on a quantification of the radiated power, current quench time and the maximum
runaway current, which must remain within defined limits for machine protection.

We find that the maximum runaway current is a non-monotonic function of the injected deuterium density, with an optimal range identified, that shows a strong sensitivity to the opacity of the plasma to Lyman radiation. The two-stage injection is found to be effective in reducing both the hot-tail seed and the transported energy losses. However, during nuclear operation, additional runaway seed sources, tritium decay and Compton scattering, make it difficult to reach tolerably low runaway currents.        

  The rest of this paper is structured as follows: in
  section~\ref{sec:model}, we briefly introduce \DREAM\ and detail the SPI model that is used in this work.
  Section~\ref{sec:performance} describes the effect of two-stage SPI
  with different injection parameters, with an emphasis on how the
  disruption dynamics is influenced by dividing the injection in two
  stages, the difference between nuclear and non-nuclear operation,
  and the importance of opacity to Lyman radiation. The influence of the model assumptions on the results presented is discussed further in section~\ref{discussion}, and conclusions are
  summarised in section~\ref{conclusions}.

\section{Shattered pellet injection and plasma dynamics}
\label{sec:model}
The evolution of the plasma parameters and runaway electrons during the disruption mitigation are modelled here with the 1D-2P
numerical tool \DREAM~\cite{DREAMPaper}. 
\DREAM\ was
extended in this work to include the capability
to model SPI based on a Neutral Gas
Shielding (NGS) model \cite{Parks_TSDW}.

The 
employed modelling accounts for the plasma response to the
SPI, transport of thermal energy by magnetic perturbations and the kinetic evolution of the electron distribution function, capturing the hot-tail as well as the Dreicer runaway generation.  
In the non-nuclear phase of tokamak operation, we find that the runaway seed is dominated by
  hot-tail generation. In the nuclear phase of operation, additional
  runaway seeds are provided by tritium decay and Compton scattering of
  gamma photons from the radioactive wall. These are modelled as quasi-stationary sources feeding particles directly into the runaway population \cite{Fulop20}. After the thermal quench, when the hot-tail mechanism is no longer active, the Dreicer mechanism is also modelled in a similar fluid-like fashion \cite{DREAMPaper}.
	
	The amplification of the runaway
  seed by the avalanche mechanism is modelled
  accounting for effects of partial screening \cite{HesslowNF}. This
  introduces a
  significant dependence of the runaway avalanche on collisional-radiative
  processes and the resulting distribution of ionization states. As we will show, the
  response to a large deuterium SPI is significantly influenced by
  the opacity of the plasma to Lyman radiation.
For convenience, an overview of the particle and energy balance, and self-consistent electric field evolution, implemented in \DREAM\ are given in \ref{app:dream}.  

An SPI starts with a pellet being accelerated and then shattered, resulting in a plume of pellet shards of different sizes and velocities entering the plasma. As the shards travel through the plasma, they act as a set of localised sources of particles. The corresponding density increment---here calculated assuming instantaneous homogenisation over the flux surfaces---depends on the ablation rate of the shards, the area of the flux surfaces, and a kernel function defining the radial spread of the deposited material around the shards.

The technical aspects needed to describe the SPI scenario are presented below in more detail. These include the size and velocity distributions of the shattered pellet shards, the explicit form of the ablation rate, and how that is translated to a density source.

 \subsection{Shattering}
We assume the pellets are shattered into $N_\mathrm{s}$ approximately spherical shards, with sizes described by an equivalent radius $r_{p,k}$, with $k=1,...,N_\mathrm{s}$. The shard radii are drawn randomly from the distribution with probability density
$	P(r_{p,k})=k_p^2r_{p,k}K_0(k_p r_{p,k})$,
where $K_0$ is the zeroth modified Bessel function of the second kind, $
k_p=\left(N_\mathrm{inj}/6\pi^2n_pN_s\right)^{-1/3}$, 
$n_p$ is the number density of the solid pellet material, and $N_\mathrm{inj}$ is the total number of injected atoms \cite{ParksGA2016}. 
This distribution of shard sizes has recently been used in several other SPI studies \cite{DiHu, AkinobuREM, NardonIAEA}. 

\begin{figure}
    \begin{center}	
    \includegraphics[width=\columnwidth]{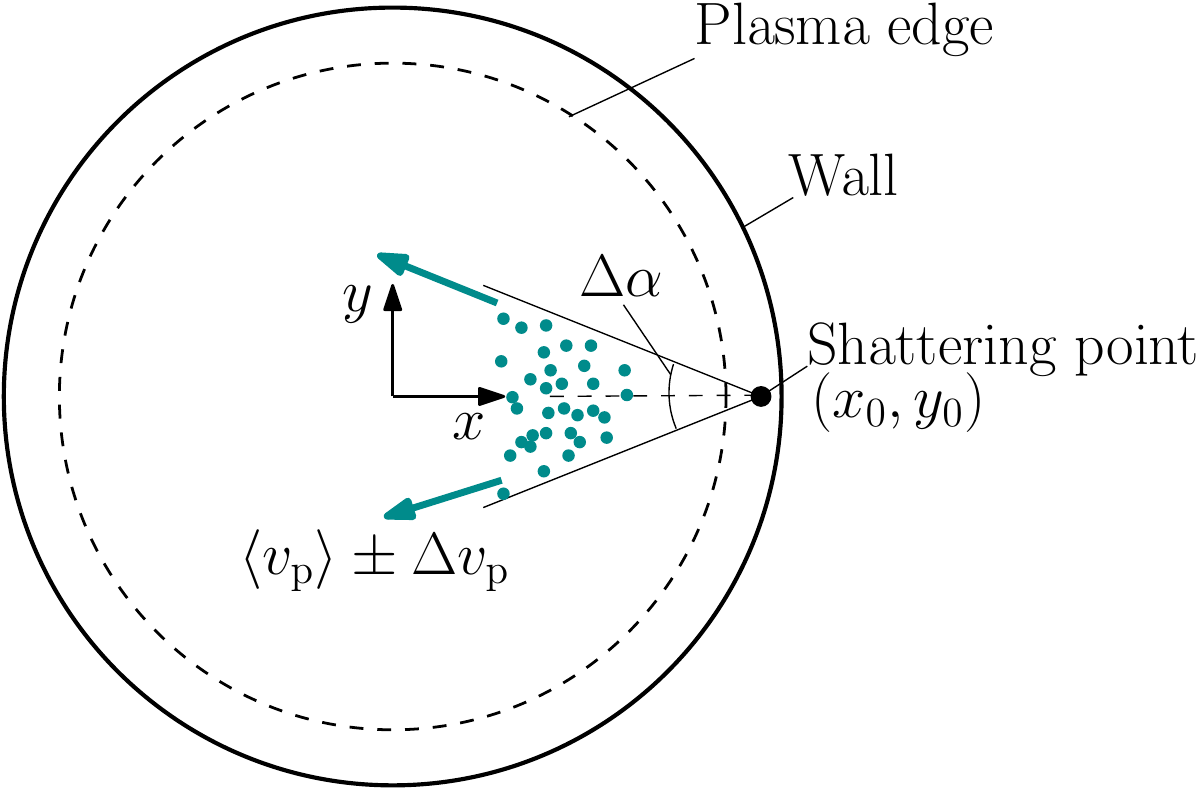}
    \end{center}
	\caption{Illustration of the modelled geometry of SPI injection. The statistics of the motion of the shards is parametrized by the mean value $\langle v_p \rangle$ and the spread $\Delta v_p$ of their speeds, as well as the divergence of the shard plume $\Delta \alpha_p$. The trajectories diverge from the shattering point $(x_0, y_0)$. }
	\label{fig:SPIillustration}
\end{figure}

Once shattered, the shards are assumed to travel with constant velocities $v_{p,k}$ in a poloidal plane, starting at the shattering point $(x_0,y_0)$, as illustrated in figure~\ref{fig:SPIillustration}; 
\begin{equation}
  \boldsymbol{x}_{p,k}(t)= (x_0-v_{p,k} \cos{\alpha _{p,k}}t, y_0+ v_{p,k} \sin{\alpha _{p,k}}t),
	\label{eq:ShardMotion}
\end{equation}
with the origin of the $(x,y)$-coordinate system taken at the plasma center. The speeds $v_{p,k}$, and angles $\alpha _{p,k}$ with respect to the horizontal plane, are chosen from uniform random distributions within $\langle v_p \rangle\pm\Delta v_p$ and $\pm\Delta \alpha _p$, respectively. The parameters $x_0$, $y_0$, $\langle v_p \rangle$, $\Delta v_p$ and $\pm\Delta \alpha _p$, as well as $N_\mathrm{inj}$ and $N_\mathrm{s}$, are considered controllable free parameters. In reality, the control of these parameters is achieved by adjusting the impact speed and angle of the pellet on the shattering surface, likely resulting in correlations between them. 

\subsection{Ablation}
We characterise the ablation rate by the time derivatives of the shard radii, $\dot{r}_{\mathrm{p},k}$. The expression used for $\dot{r}_{\mathrm{p},k}$ is based on a version~\cite{Parks_TSDW} of the NGS model~\cite{Parks_Thurnbull} that allows the pellet material to have both hydrogenic and noble gas components. Expressed in terms of the unidirectional incident heat flux $q_\mathrm{in}$ carried by the bulk plasma electrons and their effective energy $\mathcal{E}_\mathrm{in}$, this model gives
\begin{equation}
	\dot{r}_{p,k}=-\frac{\lambda (X)}{4\pi r_{p,k}^2\rho _{\rm dens}}\left( \frac{q_\mathrm{in}}{q_0}\right)^{1/3}\left(\frac{\mathcal{E}_\mathrm{in}}{\mathcal{E}_0} \right)^{7/6}\left(\frac{r_{p,k}}{r_{p0}}\right)^{4/3}.
	\label{eq:NGSextended}
\end{equation}
Here, the normalising radius, heat flux and effective energy are $r_{p0}=2$ mm, $q_\mathrm{0}=n_0\sqrt{2T_0^3/(\pi m_e)}$ and $\mathcal{E}_0=2T_0$, respectively, with the representative temperature and density $T_0=2000$ eV and $n_0=10^{20}$ m$^{-3}$. The solid mass density of the pellet is denoted $\rho _\mathrm{dens}$. The dependence of the ablation rate on the deuterium-neon composition is accounted for by the factor $\lambda(X) = [27.0837+\tan{(1.48709X)}]/1000 \;\rm kg/s,$ where $X=N_\mathrm{D_2}/(N_\mathrm{D_2}+N_\mathrm{Ne})$ is the deuterium fraction, $N_\mathrm{D_2}$ is the number of deuterium \emph{molecules} (thus the number of deuterium atoms is $N_\mathrm{D}=2N_\mathrm{D_2}$) and $N_\mathrm{Ne}$ is the number of neon atoms in the pellet. 

The heat flux and effective energy are calculated from a general electron momentum distribution function, $f$, according to
\begin{equation}
	q_\mathrm{in}= \frac{1}{4}\int m_\mathrm{e}c^2(\gamma-1)v f \,\mathrm{d}\boldsymbol{p}
	\label{eq:heatflux}
\end{equation}
and
\begin{equation}
	\mathcal{E}_\mathrm{in}= \frac{2}{n_\mathrm{free}}\int m_\mathrm{e}c^2(\gamma-1) f \,\mathrm{d}\boldsymbol{p}.
	\label{eq:E0}
\end{equation}
We note that equation~\eqref{eq:NGSextended} was derived assuming a Maxwellian electron momentum distribution, with temperature $T_M$. However, it may be assumed to be sufficiently accurate for the small deviations from a Maxwellian present in the early stages of the disruption while the shards are still ablating, that is, before a substantial runaway acceleration has occurred\footnote{In the presence of non-negligible runaway populations different physics mechanisms dominate \cite{Kiramov_2020}.}. The total free electron density is $n_\mathrm{free}=\int f\mathrm{d}\boldsymbol{p}$, $c$ is the speed of light, and $\gamma$ is the Lorentz factor. The factor $1/4$ in equation~\eqref{eq:heatflux} converts the isotropic heat flux to the average unidirectional heat flux facing the pellet shards, and is strictly valid for a Maxwellian distribution \cite{stangeby2000plasma}. The normalisation of $\mathcal{E}_\mathrm{in}$ is chosen such that $\mathcal{E}_\mathrm{in}$ reduces to $2T_\mathrm{M}$ for completely Maxwellian electrons, which is equal to the ratio of the unidirectional heat flux and the unidirectional particle flux. 

\subsection{Material deposition}
The ablated material quickly ionizes, becoming confined by the magnetic field to the flux surfaces near the pellet shard position where the ablation took place. 
The homogenization and equilibration of the ablated material is approximated here to take place instantaneously, an assumption also made in other recent SPI studies \cite{Nardon_2020, AkinobuREM, NardonIAEA, ShirakiIAEA}. The impact of this assumption is discussed later in this paper. 

The homogenized ion density increase on the flux surface with radius $r$ 
is given by
\begin{equation}
	\left(\frac{\partial n_{ij}}{\partial t}\right)_\mathrm{SPI}=-f_{ij} \sum _{k=1}^{N_\mathrm{s}} \frac{4\pi r_{p,k}^2\dot{r}_{p,k}\rho _{\rm dens}N_\mathrm{A}}{\mathcal{M}}H(r, \rho _{p,k}),
\label{eq:deposition}
\end{equation}
where the pellet molar mass is  $\mathcal{M}$, and $N_\mathrm{A}$ is the Avogadro number. A fraction $f_{ij}$ of the ablated pellet material appears in the charge state $i$ of ion species $j$, with a corresponding density $n_{ij}$. 

The radial distribution of the deposited material, in terms of density increase, is described by the factor $H(r, \rho _{p,k})=h(r,\rho _{p,k})/A_\mathrm{fls}(r)$, where $A_\mathrm{fls}=4\pi ^2 rR_0$ is the area of the flux surface at radius $r$ and $h(r,\rho _{p,k})\mathrm{d}r$ describes the fraction of the material deposited at a radius between $r$ and $r+\mathrm{d}r$ ablated from a pellet located at minor radius $\rho _{p,k}$. 
The width of this \emph{deposition kernel} is physically determined by transport processes radially dispersing the ablated material. Some previous studies have used a Gaussian deposition kernel $h\propto \exp{[(r-\rho _{p,k})^2/r_\mathrm{cld}^2]}$ \cite{DiHu, Nardon_2020}, with a "cloud width" of $r_\mathrm{cld}$ that has a lower bound of $\sim 1\,\rm cm$ \cite{Lengyel_1999}, the width of the flux tube channeling the pellet material in the vicinity of the shard. After verifying that $r_\mathrm{cld}$ has only a weak quantitative impact on the final density profile, we opted for a delta function deposition kernel, $h=\delta (r-\rho _{p,k})$ for numerical convenience, translating to a uniform distribution over the distance travelled by the shard during one time step.

To avoid the need to resolve the extremely rapid ionization dynamics of the ablated material, we assume the deposited material  appears in its equilibrium distribution of charge states, at the local density and temperature. With $\phi _j$ denoting the particle fraction of the pellet material consisting of species $j$, the quantity $f_{ij}$ appearing in equation~\eqref{eq:deposition} becomes $f_{ij}=\phi _jn_{ij}^\mathrm{eq}/n_{\mathrm{tot},j}$, where the equilibrium distribution of charge states is calculated according to
\begin{eqnarray}
    &R_{ij} n_{i+1,j}^\mathrm{eq}-I_{ij} n_{ij}^\mathrm{eq}=0,  \hspace{0.5cm}i=0,1,...,Z-1,\\
    &\sum_i n_{ij}^\mathrm{eq}=n_{\mathrm{tot},j},  \hspace{0.5cm}i=0,1,...Z.
     \label{eq:ioniz_equilibrium}
\end{eqnarray}
The total density of species $j$, with atomic number $Z$, is denoted by $n_{\mathrm{tot},j}$, and $I_{ij}(T_\mathrm{M},n_\mathrm{M})$ and $R_{ij}(T_\mathrm{M},n_\mathrm{M})$ are the ionization and recombination rates, respectively, obtained from the OpenADAS database \cite{ADAS}. 

The energy required for the initial ionization is accounted for by a loss term \eqref{eq:additionalionizationloss} in the energy transport equation~\eqref{eq:heateq}.

\subsection{Plasma response to SPI}
\DREAM\ calculates the time evolution of the pellet and plasma parameters self-consistently, within a model summarised as follows. The pellet shards follow straight lines as described by equation~\eqref{eq:ShardMotion}, and the evolution of the shard sizes is governed by the ablation rate, equation~\eqref{eq:NGSextended}. The resulting density increase is given by equation~\eqref{eq:deposition}, and the resulting cooling is calculated by the heat transport equation~\eqref{eq:heateq} with the additional loss term given in equation~\eqref{eq:additionalionizationloss}.

Once the pellet material is deposited, the evolution of $n_{ij}$ is
governed by ionization and recombination according to the \emph{time
dependent} rate equations~\eqref{eq:tdre}. The electric field evolution resulting from the rapid change in the conductivity is given by the
induction/diffusion process \eqref{eq:ampere}. The momentum distribution of the
electrons is resolved using the ``fully kinetic'' mode described
in~\cite{DREAMPaper}.

\section{Simulation of two-stage shattered pellet injection}
\label{sec:performance}
In this section we evaluate the disruption mitigation performance of the two-stage deuterium-neon injection scheme in an ITER-like plasma.
With this scheme, the plasma is first cooled by dilution down to the $100-1000$ eV range by a deuterium injection, without significantly affecting the thermal energy density (and hence the pressure). A few $\rm ms$ later, neon is injected, with the aim of radiating away most of the thermal energy, and producing a current quench with an acceptable timescale. The temporal separation between the injections allows the tail of the electron distribution to thermalize, thereby reducing hot-tail runaway generation. Another advantage is that the ratio of the transported and radiative energy losses decreases if the thermal quench is triggered from a lower temperature, reducing the risk of damage in plasma facing components due to excessive localized heat loads. These aspects are therefore emphasised in our investigation.

\subsection{Injection and plasma parameters}
\label{sec:params}
We search for suitable injection parameters for ITER-like plasmas with initial temperature profile $T_\mathrm{M}(r) = T_c(1-0.99(r/a)^2)$, central temperature $T_c=20$ keV, and a radially constant initial density $n_\mathrm{M}=10^{20}$ m$^{-3}$. The profiles used here have previously been used to study massive material injections in ITER-like scenarios \cite{ITER_OV,MS2017}, assuming flat deposition profiles. For the initial ion composition, we consider both a pure deuterium plasma and an even mix of deuterium and tritium, the latter corresponding to the nuclear operation phase. The plasma minor radius is $a=2$ m, the radius of the perfectly conducting wall 
is $b=2.15$ m and the major radius of the tokamak is $R_0=6.2$ m. The initial plasma current is $I_\mathrm{p}=15$ MA, with radial profile given by $j_{||}(r)=j_0(1-(r/a)^2)^{0.41}$, with $j_0=1.69$ MA/m$^2$. 

A fast reaction time of the disruption mitigation system favours high pellet speeds. We therefore consider the fastest injection speeds expected. These are $\langle v_\mathrm{p,D} \rangle = 800$ m/s for deuterium pellets and $\langle v_\mathrm{p,Ne} \rangle = 200$ m/s for neon pellets \cite{LehnenIAEA}. For the distribution of pellet speeds we assume $\Delta v_\mathrm{p}=0.2 \langle v_\mathrm{p} \rangle$ and we set the divergence angle $\Delta \alpha _\mathrm{p}=20^\circ$, for both deuterium and neon pellets. This angle primarily affects how many shards pass through the innermost flux surfaces, and an increased divergence can shift the deposited density profile from the core further out in the plasma. While this may be of interest when fine-tuning the density profile, we are able to achieve satisfactory density profiles with this fixed value. The current quench dynamics is expected to be rather insensitive to the details of the density profile for a given number of injected particles, as long as core penetration is achieved~\cite{ITER_OV}. We also fix the position at which the pellets are shattered to the tokamak wall on the horizontal mid-plane, i.e.~at position $(x_0,y_0)=(b,0)$.

The remaining injection parameters to investigate are the number of injected deuterium and neon particles, $N_\mathrm{inj,D}$ and $N_\mathrm{inj,Ne}$, and the number of shards, $N_\mathrm{s,D}$ and $N_\mathrm{s,Ne}$. In order to make the most efficient use of the pellet material, the number of shards for a given number of injected particles should be chosen to achieve core penetration without leaving unablated pellet material. To this end, we perform a scan of the assimilation rate, i.e.~the fraction of the pellet material that is deposited in the plasma, as a function of injected quantities and number of shards.

\begin{figure}
  \centering
    \includegraphics[width=1.0\columnwidth]{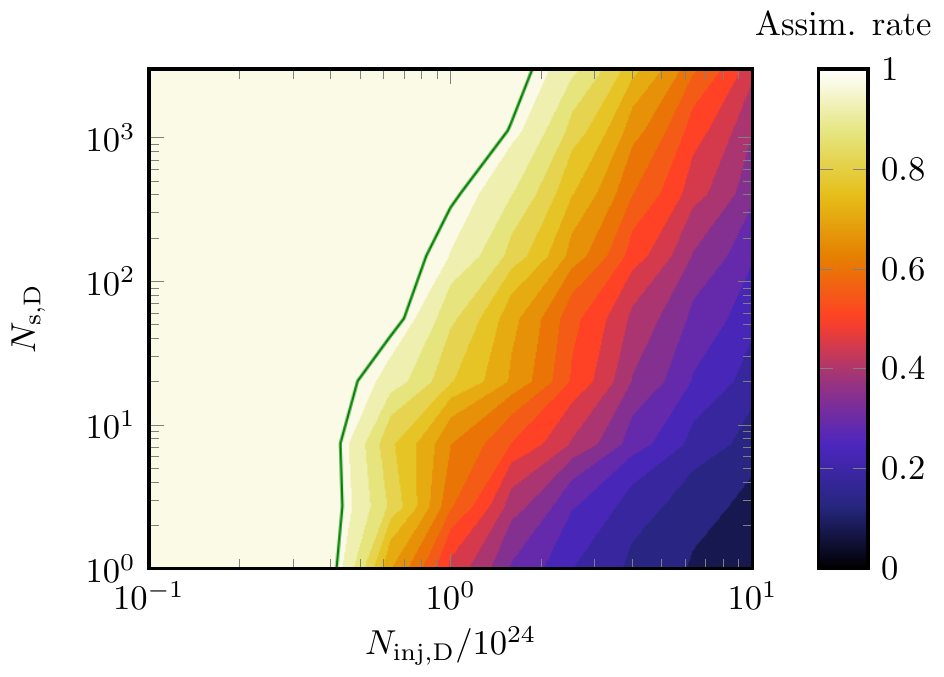}
  \caption{Assimilation rate as a function of the number of injected deuterium atoms $N_\mathrm{inj,D}$ and number of shards $N_\mathrm{s,D}$.   The green line marks the 97\% assimilation contour, along which $N_\mathrm{s,D}$ and $N_\mathrm{inj,D}$ are chosen in the later simulations in this paper.}
  \label{fig:SPI_D}
\end{figure}

The deuterium assimilation rate as a function of $N_\mathrm{inj,D}$ and $N_\mathrm{s,D}$ is shown in figure~\ref{fig:SPI_D}. For a given $N_\mathrm{inj,D}$ the optimal choice of $N_\mathrm{s,D}$ is thus close to the 100\% assimilation contour. As we will show later, having a good core penetration is essential for a successful disruption mitigation. 
To ensure a sufficient margin, the number of injected deuterium particles and shards in this paper are chosen along the 97\% contour marked by the green line in figure~\ref{fig:SPI_D}. 

The relatively low temperature of the diluted plasma when the neon pellets are injected makes their full assimilation difficult. Our studies indicate that the increasing trend in the assimilation rate with increasing number of shards slows down at $N_\mathrm{s,Ne}\sim 100$. Therefore, we fix the number of neon shards to 100.

\subsection{Representative disruption dynamics}
\begin{figure*}
	    \includegraphics[width=1\textwidth]{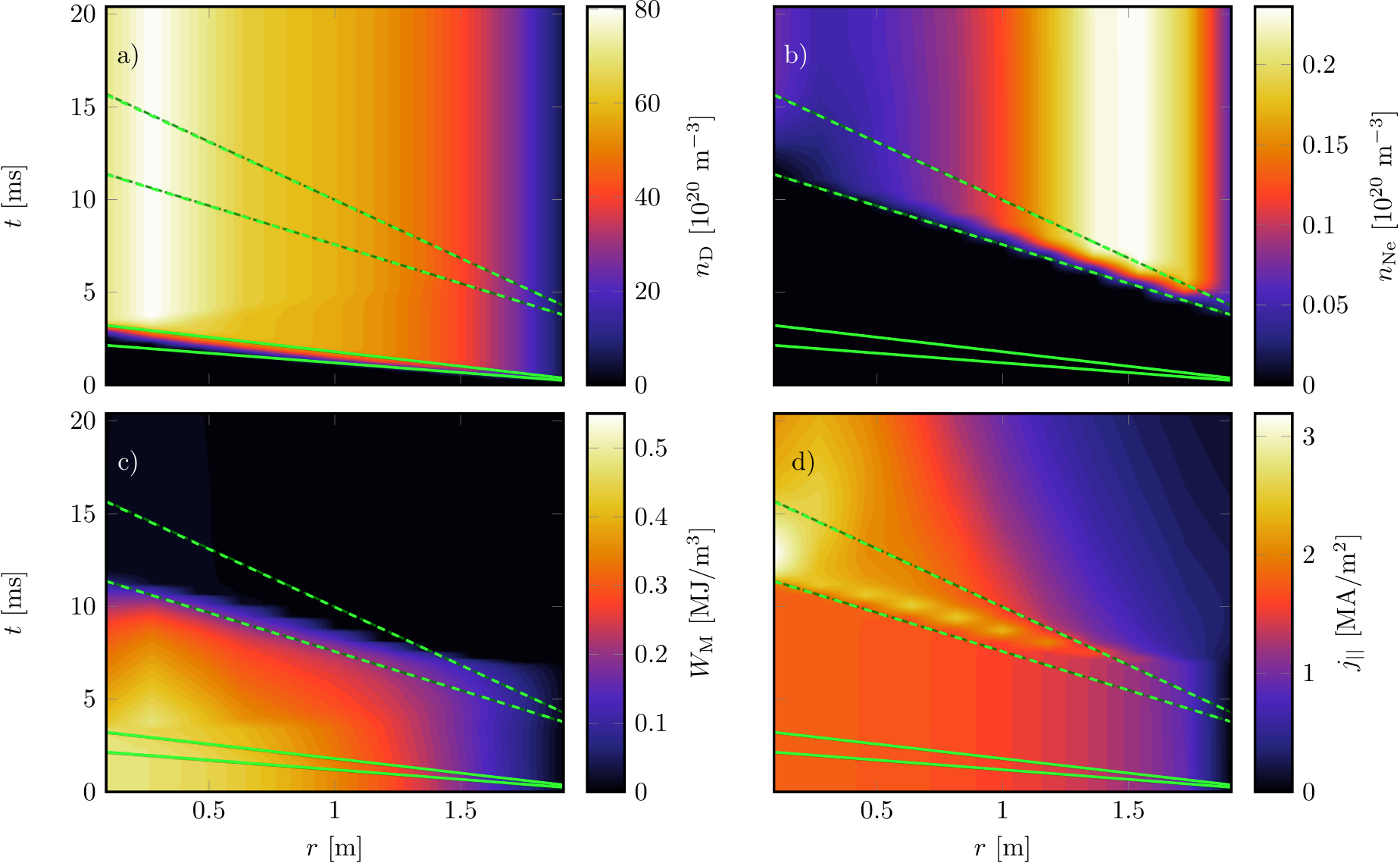}
	\caption{Spatio-temporal evolution of the deuterium density (a), neon density (b), electron thermal energy density (c), and current density (d) during a two-stage SPI injection with parameters $N_\mathrm{s,D}=1688$, $N_\mathrm{inj,D}=2\cdot 10^{24}$, $N_\mathrm{s,Ne}=100$, and $N_\mathrm{inj,Ne}=1\cdot 10^{23}$. The mean speed of the deuterium shards is $\langle v_\mathrm{p,D} \rangle=800\,\rm m/s$ and of the neon shards is $\langle v_\mathrm{p,Ne} \rangle=200\,\rm m/s$. A diffusive heat transport corresponding to $\delta B/B=7\cdot 10^{-4}$, constant in space and time, is activated once the neon shards enter the plasma. }
	\label{ht_case_study_2_double}
\end{figure*}
We now consider the spatio-temporal evolution of the most relevant plasma parameters during a representative two-stage injection, using the parameters $N_\mathrm{inj,D}=2\cdot 10^{24}$\footnote{This quantity roughly corresponds to a $28\,\rm mm$ pellet of the ITER disruption mitigation system.}, $N_\mathrm{s,D}=1688$, $N_\mathrm{inj,Ne}=10^{23}$ and $N_\mathrm{s,Ne}=100$. The neon shards are injected at $t=3.4\,\rm ms$, when the last deuterium shards have just left the plasma centre.  We model the diffusive radial electron heat transport  due to magnetic perturbations. We switch on a perturbation amplitude of $\delta B/B=7\cdot 10^{-4}$ when the neon shards---that create significant pressure variation---enter the plasma, giving a characteristic time scale of the heat transport of about 1 ms. 
As the purpose of this study is to investigate trends of disruption mitigation performance measures using a two-stage injection, we consider a simplified case without radial transport of superthermal electrons, reducing the runtime by about an order of magnitude. Such a conservative case can be regarded as an upper limit of the runaway seed generation during the thermal quench, primarily by the hot-tail mechanism. The consideration of such a case is further motivated by the lack of detailed knowledge about the transport of superthermals. The final runaway current has also previously been found to be a logarithmically weak function of the runaway seed during ITER-like conditions \cite{ITER_OV}, and therefore the omission of radial transport of superthermal electrons is not expected to substantially alter the trends found in this study.

The spatio-temporal evolution of the deuterium and neon densities, the electron thermal energy and the current density are shown in figure~\ref{ht_case_study_2_double}. The two solid green lines mark the trace of the fastest and slowest deuterium shards (for shards with $\alpha _\mathrm{p}=0$), and the dashed green lines mark the corresponding traces for the neon shards. In figure~\ref{ht_case_study_2_double}a-b, we see a clear increase in the deuterium density and neon density between the corresponding pair of green lines. Note the increase in the neon density saturates quite closely to the first dashed green line, especially in the inner part of the plasma. The neon deposited by the first shards radiatively cools the plasma, impeding the ablation of the later shards. 

The sudden deposition of the released thermal energy content during the thermal quench might cause melting of plasma facing components if the heat loads are localised. It is therefore necessary for the disruption mitigation system to ensure that a major part of the thermal energy is lost through radiation. In ITER, the radiated fraction should be larger than 90\% of the initial thermal energy content.  Looking at figure~\ref{ht_case_study_2_double}c, we see that the thermal energy density is only slightly affected by the deuterium injection. The plasma is cooled mainly by dilution in this phase, by a factor corresponding to the density increase, resulting in temperatures of a few hundred $\rm eV$. When the neon shards enter the plasma, the thermal energy density is dissipated---and the temperature drops---over a millisecond time scale or faster in the parts of the plasma that have been reached by the neon shards. The onset of the magnetic perturbations causes some diffusion of the thermal energy density in the parts of the plasma that have not yet been reached by the neon shards. However, in total, diffusive transport dissipates only 7\% of the initial thermal energy. Notably, this value is smaller than the 10\% limit, and much smaller than the radiative losses that dissipate almost all of the remaining thermal energy. Although it should be emphasised that our model for the thermal energy diffusion is rather crude, assuming here an immediate onset of a prescribed magnetic perturbation that then remains constant in time and space, it allows a simple comparison of the calculated transported losses in different scenarios and can readily be improved for more detailed studies.

The temperature drop to a few $\rm eV$ results in the onset of the current density drop in the part of the plasma which has been reached by the neon shards, as seen in figure~\ref{ht_case_study_2_double}d. We also see a radial spike in the current profile moving inwards along with the neon shards. This spike is caused by the diffusion of the electric field, induced where the plasma has been cooled, into the hotter region that the neon shards have not yet reached. In this region, the conductivity is still high enough that even a relatively modest increase of the electric field can cause a significant increase in the (ohmic) current density. When the neon shards have reached the core, in most cases, the current starts to decay in all parts of the plasma. In some cases the ohmic heating, amplified by the radial current spike, can cause parts of the plasma to re-heat. The re-heating is accompanied by an increase in the conductivity and, as the electric field diffuses into the re-heated regions, the current density can initially increase locally. The decay of the total current is comparatively slow, with a time scale of the order of seconds.

\subsection{Radiative vs transport losses}
As the radiative loss fraction in a disruption is required to be large, we studied the sensitivity of the fraction of the thermal energy lost by transport during two-stage injections to the number of injected neon and deuterium particles. In addition to pellet parameters chosen as described in section~\ref{sec:params}, we illustrate the consequences of not achieving core penetration, by including simulations of smaller deuterium pellets which are fully ablated before they reach the core.
For these pellets, we chose $N_\mathrm{s,D}=10$. 

\begin{figure}
  	\centering
  	    \includegraphics[width=1.0\columnwidth]{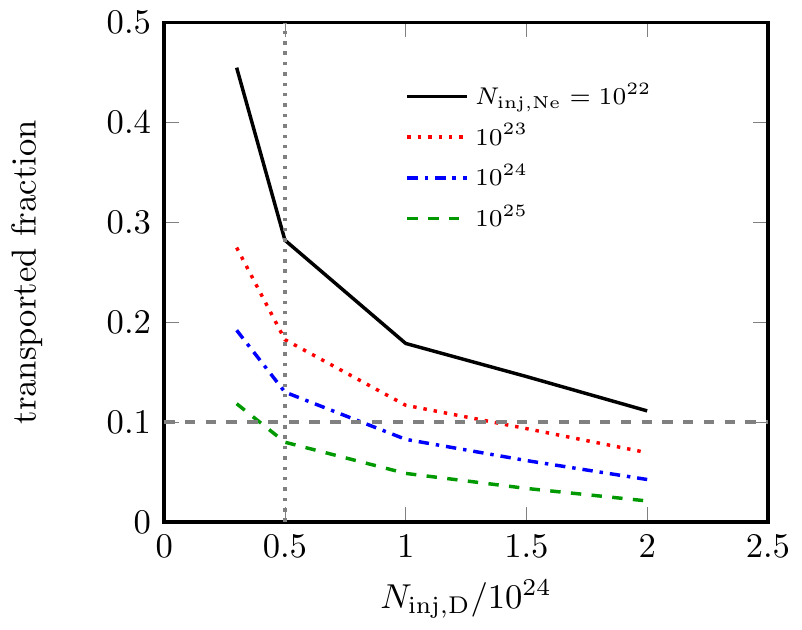}	
	\caption{Fraction of the thermal energy lost by transport due to magnetic field perturbations, as a function of the number of injected deuterium and neon particles. The parameters for the injections were $N_\mathrm{s,Ne}=100$ for the neon shards, and the number of deuterium shards was chosen along the assimilation contour in figure~\ref{fig:SPI_D}, except to the left of the dotted grey line, where core penetration is not achieved, and we set $N_\mathrm{s,D}=10$. A diffusive heat transport corresponding to $\delta B/B=7\cdot 10^{-4}$, constant in space and time, is activated once the neon shards enter the plasma. Dashed grey line marks the target of a transported fraction lower than 10\%.}
	\label{fig:transp_frac}
\end{figure}

Figure~\ref{fig:transp_frac} shows the fraction of the initial thermal energy that is lost by transport as a function of the number of injected deuterium and neon particles. We see that the transported fraction can be significantly decreased by increasing the deuterium and neon content. The dependence on $N_\mathrm{inj,Ne}$ is weaker than the dependence on $N_\mathrm{inj,D}$, mainly due to the decrease in the assimilated fraction as $N_\mathrm{inj,Ne}$ increases. The neon content primarily increases the radiative losses, which are further enhanced by the increase in the electron density due to the deuterium injection. 
The deuterium content also limits the transport by lowering the temperature before the onset of the magnetic perturbation. It is noteworthy that simulated transported losses below the limit of 10\% of the initial thermal energy, marked by the dashed horizontal line in figure~\ref{fig:transp_frac}, are achievable within a realistic range of injection parameters. Again, further studies with alternative perturbation configurations will be needed to determine the robustness of the observed trend of reduced transported energy losses during two-stage deuterium-neon SPI.  

Finally, we note that lack of core penetration does not cause a dramatic difference in the transported fraction. One reason for this is that the neon shards can reach the core even if the deuterium shards did not. Another reason is that the thermal energy in the core has to pass the outer regions of the plasma before it can be lost due to transport, and can therefore be radiated away in the regions that have been reached by the deuterium shards.

\subsection{Current decay and runaway generation}
We now turn our attention to the later stages of the disruption and study the decay of the ohmic current, as well as the runaway generation and dissipation.
At this stage, the thermal energy is almost completely dissipated, and the already low plasma temperature evolves slowly as the ohmic heating varies.
When the rapid plasma cooling is complete, flux surfaces are expected to re-heal, which we account for by switching off the transport due to magnetic perturbations. The duration of the current quench must be limited to around $50-150\,\rm  ms$ to limit the forces experienced by the vessel. 

We consider separately the non-activated and the nuclear operation phases, with the tritium decay and Compton scattering seed mechanisms being active in the latter. These mechanisms generate a seed runaway current of the order of $0.1-1\,\rm A$ rather independently of the injection parameters. These seeds are sufficiently large to be multiplied to a final runaway current of several MA by the avalanche mechanism, as we will show.

In the relatively cold plasma undergoing a current quench, the radiation transport properties of the plasma can significantly impact the runaway generation. In particular, a preliminary estimate of opacity to Lyman radiation was presented in reference~\cite{ITER_OV}, indicating that this effect could lower the runaway currents by up to several MA for large injected deuterium densities. Not only does the opacity impact the radiative losses, but it also affects the ionization and recombination rates \cite{Pshenov_2019}. While the plasma remains essentially transparent at most wavelengths, the opacity is significantly increased at the wavelength corresponding to resonant transitions \cite{Morozov}. This applies particularly to those transitions involving the ground state; as this is the state occupied by most ions and atoms.
The estimate shown in \ref{app:opacity} indicates that the plasma may only be transparent to a few percent of the Lyman radiation resulting from excitations from the ground state, which mostly populates the lower excited states.
This impacts the contribution from hydrogen isotopes to the radiation rate $L_\mathrm{ij}$. On the other hand, the plasma is estimated to be transparent to the majority of the recombination radiation. A substantial part of this radiation consists of a continuum spectrum resulting from free-bound transitions, together with higher order Lyman lines resulting from the de-excitation of the high excited states thus populated. 
Opacity to neon radiation is not expected to have a strong impact on disruption dynamics \cite{Lukash}. We therefore consider the limiting cases where the plasma is assumed to be completely transparent or completely opaque to Lyman radiation, whilst remaining completely transparent to radiation from species other than hydrogen. For the completely transparent case, the radiation and ionization/recombination rates are taken from the ADAS database for all species. When the plasma is assumed to be opaque to Lyman radiation, these data are instead taken from the AMJUEL database\footnote{http://www.eirene.de/html/amjuel.html}. 

\begin{figure*}
	\centering
	    \includegraphics[width=0.9\textwidth]{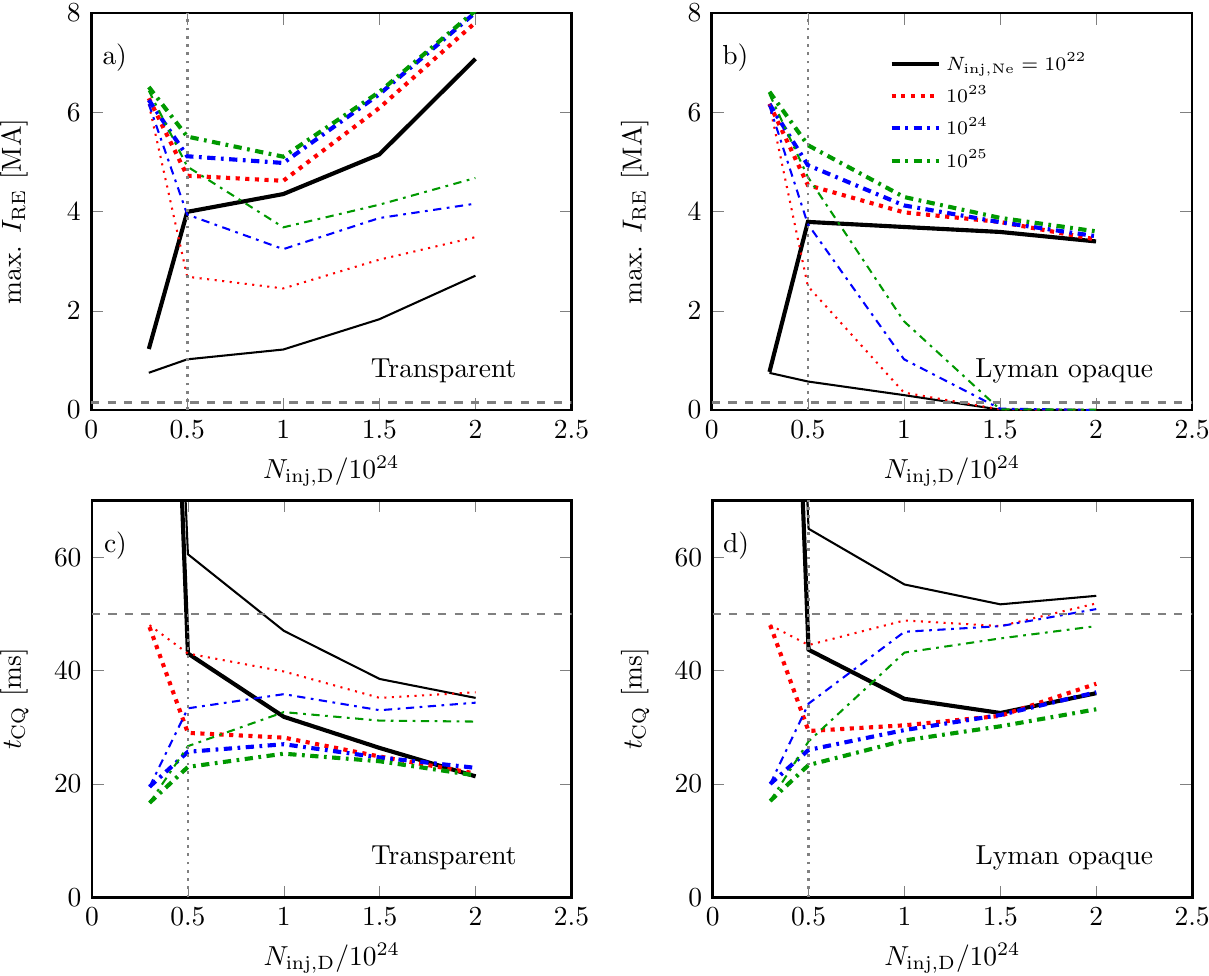}
	\caption{Maximum runaway current (a-b) and current quench time (c-d) as functions of the number of injected deuterium and neon atoms. Thick lines correspond to nuclear operation, with tritium decay and Compton scattering runaway sources, thin lines correspond to non-activated operation. In panel a) and c), the plasma is completely transparent, while it is opaque to the Lyman lines in panels b) and d). The injection parameters are $N_\mathrm{s,Ne}=100$, with $N_\mathrm{s,D}$ chosen along the assimilation contour in figure~\ref{fig:SPI_D}, except to the left of the dotted grey line, where $N_\mathrm{s,D}=10$. Electron heat transport corresponding to a magnetic perturbation of $\delta B/B=7\cdot 10^{-4}$ is activated when the neon shards enter the plasma, and switched off at $t=16.4$ ms, when the rapid plasma cooling is complete.}
	\label{fig:maxRE}
\end{figure*}

Figures~\ref{fig:maxRE}a-b show the maximum of the runaway current for the different scenarios discussed above, indicating an upper bound of the runaway current that may strike the wall. The thin lines show the non-nuclear case, while the thick lines show results for the nuclear case, i.e.~with the tritium decay and Compton scattering seed mechanisms included. Figures~\ref{fig:maxRE}c-d show the corresponding ohmic current quench times, defined as 
\begin{equation}
t_{\rm CQ}=\frac{t(I_{\rm Ohm}=0.2I_{\rm p}^{(t=0)})-t(I_{\rm Ohm}=0.8I_{\rm p}^{(t=0)} )}{0.6}.
\end{equation}
In figure~\ref{fig:maxRE}a and c the plasma is assumed to be completely transparent, and in figure~\ref{fig:maxRE}b and d it is completely opaque to Lyman radiation.

The first observation in figure~\ref{fig:maxRE}a-b concerns the cases with the lowest considered $N_\mathrm{inj,D}$ and $N_\mathrm{inj,Ne}$ values (leftmost points of solid curves), where core penetration of deuterium is not achieved. In these cases the radiative cooling is not strong enough to overcome the ohmic heating. This leads to a re-heating of the plasma from the $10\,\rm eV$ range to a few hundred $\rm eV$, once the transport losses are no longer active. The re-heating greatly increases the conductivity, leading to a major reduction of the current decay rate, as well as the induced electric field and the runaway generation rate. In these cases the runaway currents remain relatively small, however the current quench times are unacceptably long (outside the plotted range in panels c-d). The current quench was not completed within the  $150\,\rm ms$ of the simulation, but the decay rates indicate a current quench time scale of seconds.

In all other cases shown in figure~\ref{fig:maxRE}c-d, the current quench times are in the vicinity of the lower acceptable limit of 50 ms, marked by dashed grey lines. Note, however, that in cases with a large runaway conversion, the conversion to runaway current aborts the current quench rather abruptly. The ohmic current quench time calculated here is therefore a lower estimate. 

Apart from these, somewhat singular, cases with the lowest injected quantities, the general trend of the maximum runaway current with an increasing $N_\mathrm{inj,D}$ is either non-monotonic---a decreasing trend turning into an increasing one---in case of a transparent plasma, or a monotonic decrease in case of a plasma opaque to Lyman radiation\footnote{Note this the non-monotonic behaviour also appears in a plasma opaque to Lyman radiation, just at higher injected deuterium quantities, outside the range shown here.}. The decreasing trend is caused by the following effects. At the post-thermal quench temperatures of a few $\rm eV$, neon is not fully ionized, while deuterium remains practically fully ionized until the temperature drops below $\sim 2\,\rm eV$. Thus, the injected deuterium contributes with a large increase in the electron density, increasing the critical electric field for runaway generation. Moreover, since a high fraction of bound electrons favours avalanche generation, and that the deuterium content decreases the fraction of bound electrons, we find that avalanche is suppressed when the injected deuterium quantity is increased.  

The non-monotonic trend observed in figure~\ref{fig:maxRE}a is explained as follows. As the ohmic current decays and the ohmic heating decreases,  the temperature eventually falls below $2\,\rm eV$ and deuterium starts to recombine.
This leads to an increase of the fraction of bound electrons, enhancing the avalanche. At high deuterium densities, the increased radiative losses can cause this to happen, already when there is still a substantial part of the ohmic current left that can be converted to runaways. How much ohmic current remains when the deuterium starts to recombine depends on the radiation transport properties of the plasma. Deuterium recombination also contributes significantly to the radiation, thus opacity to Lyman radiation can reduce the radiative losses considerably. As a consequence, with opacity the ohmic current is smaller when deuterium recombines, hence the lower maximum runaway currents at high deuterium densities in figure~\ref{fig:maxRE}b. The effect of the radiative properties of the plasma are explored further in the next subsection.

Even though a sufficiently large deuterium injection strongly reduces the hot-tail seed, when tritium decay and Compton scattering is included (thick lines in figure~\ref{fig:maxRE}a-b) the maximum runaway can not be reduced much below $4\,\rm MA$ for any combination of injected quantities (apart from the case with the unacceptably long $t_{\rm CQ}$) even in the presence of opacity effects.  
This is due to the several orders of magnitude difference between the $0.1-1\,\rm  A$ seed produced by tritium decay and Compton scattering, and the hot-tail seed. The finite runaway current depends only logarithmically on the seed, as found in reference~\cite{ITER_OV}, due to self-regulating interaction between the runaway current and the electric field. When the runaway current becomes comparable to the remaining ohmic current, the induced electric field decreases, reducing the avalanche growth rate. 
\subsection{Opacity to Lyman radiation}
\begin{figure}[H]
	\centering
	    \includegraphics[width=0.9\columnwidth]{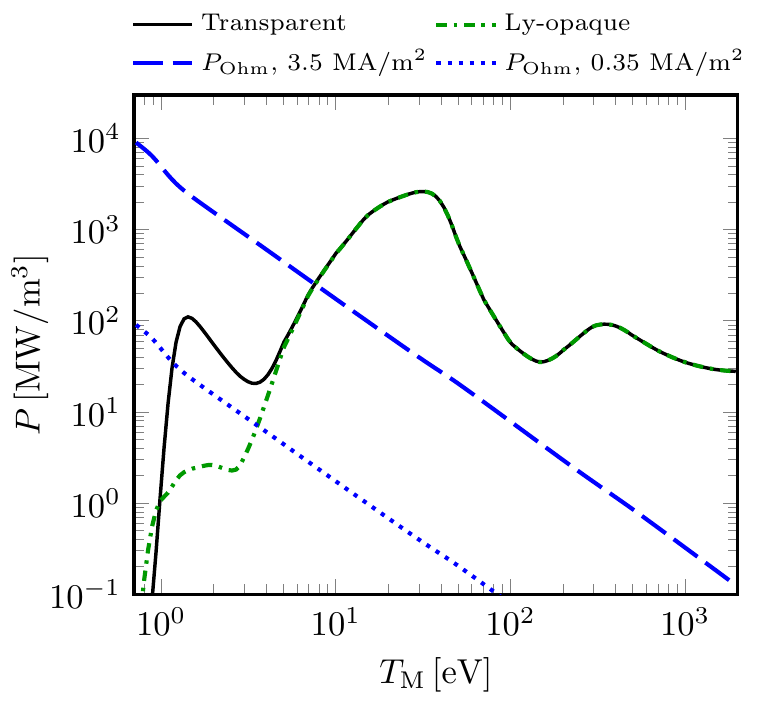}	
	\caption{Radiative power loss as a function of temperature, for a transparent plasma (solid black) and for a plasma opaque to Lyman radiation (dash-dotted green), compared to the ohmic heating calculated for $j_\mathrm{Ohm}=3.5\,\rm  MA/m^2$ (dashed blue) and $j_\mathrm{Ohm}=0.35\,\rm MA/m^2$ (dotted blue). The radiated power is calculated using the volume averaged final deuterium and neon densities of an injection with $N_\mathrm{D,inj}=2\cdot 10^{24}$ and $N_\mathrm{Ne,inj}=10^{23}$, and an equilibrium distribution of charge states.}
	\label{fig:ebalance}
\end{figure}

The differences in the dynamics between the transparent and Lyman opaque case are primarily caused by differences in the balance between ohmic heating and radiative losses during the current quench. This balance can be understood by comparing ohmic heating to radiative losses at different temperatures, assuming an equilibrium distribution of charge states (which may be calculated using equation~\eqref{eq:ioniz_equilibrium}).  
The relevant terms, corresponding to the volume averaged final densities of an injection  with $N_\mathrm{D,inj}=2\cdot 10^{24}$ and $N_\mathrm{Ne,inj}=10^{23}$, are plotted in figure~\ref{fig:ebalance}. The solid black line shows the radiative losses assuming a completely transparent plasma. The green dashed line shows the corresponding radiative losses when the plasma is assumed to be opaque to Lyman radiation. The ohmic heating is calculated for two different current densities: $3.5\,\rm MA/m^2$ (blue dashed), taken as a representative value for the maximum current density, comparable to the peak value seen in figure~\ref{ht_case_study_2_double}f, and $0.35\,\rm MA/m^2$ (blue dotted), representing the phase where the ohmic current has partially decayed.

High deuterium density corresponds to strong radiative losses at low temperatures, especially below $\sim 2\,\rm eV$, where a substantial fraction of the deuterium recombines. Through recombination, deuterium directly contributes to  radiation losses, rather than merely increasing the electron density. In a transparent plasma with $j_\mathrm{Ohm}=0.35\,\rm MA/m^2$, the equilibrium temperature is close to $1\,\rm eV$, with the ionized fraction of deuterium being only a few percent. This situation favours avalanche, as well as larger induced electric fields (albeit the field decays more rapidly). Moreover, although the ohmic current density is modest in the $\sim 1\,\rm eV$ regions, the conversion from ohmic to runaway current in a given radial location is not limited by the local ohmic current density. The current from other parts of the plasma can diffuse into the cold regions, potentially causing a local increase in the current density. These effects combined give rise to the increase in the maximum runaway current at large deuterium densities seen in figure~\ref{fig:maxRE}a. 

If the plasma is opaque to Lyman radiation, however, the radiative losses at low temperatures are considerably reduced (compare dash-dotted to solid line in figure~\ref{fig:ebalance}). This is particularly true for the line radiation following excitations from the ground state. The remaining increase in the radiative losses below $\sim 2\,\rm  eV$ is instead primarily due to recombination radiation of hydrogen species. The drop to $\sim 1$ eV is now postponed until the ohmic current has decreased further than in the transparent case. The postponed temperature drop to $\sim 1\,\rm eV$  reduces the maximum runaway current at large deuterium densities, and shifts the trend of increasing runaway current to even higher deuterium densities.

\section{Discussion}
\label{discussion}
The SPI model used in this work contains a number of simplifications. One such simplification is the use of the NGS ablation model, which is based on a simplified spherical pellet shard geometry and neglecting the details of the ablating electron momentum distribution. In addition, it neglects the electrostatic shielding~\cite{Pegourie_2004} resulting from the difference in ion and electron mobility and the magnetic shielding~\cite{Gilliard_1980} due to the deflection of magnetic field lines around the ablation cloud. Nevertheless, the NGS model has been shown to compare reasonably well with experiments, and the effect of the above simplifications have been estimated in the literature to counteract each other \cite{Pegourie_2007}. Therefore, while a more advanced ablation model could result in an order unity correction to the deposition profile, the discrepancy compared to the NGS model is expected to be quite modest. 

A larger correction might be caused by the details of the expansion process of the pellet material between the ablation and final deposition. In the present model, this process has simply been assumed to be instantaneous and local. In reality, however, the homogenisation and equilibration process takes of the order of $1\,\rm ms$. This time scale is similar to the time it takes the plume of shards to pass a given flux surface. As the newly ablated material is subject to an $E\times B$-drift towards the low field side, the deposited material will cover a wider radial region. (The drift may be less pronounced for neon than for deuterium pellets. The drive of the drift -- the excess pressure of the pellet cloud -- is directly reduced by the lower temperature of the cloud due to a radiative cooling of neon. In addition, in this two-stage injection scheme the background temperature is lower when the neon shards enter the plasma, with a correspondingly low ablation rate, leading to a lower excess density in the cloud \cite{Rozhansky_2004}.)

Apart from the direct impact on the timing and position of the density increase, the details of the expansion process might significantly alter the ablation itself. For pellets with a significant impact on the plasma temperature, the interaction between the plasma and the pellet material is self-regulating; when the pellet material is ablated, the plasma is cooled, at first primarily by dilution, resulting in a slowing down of the ablation. The finite expansion time and drifts alter this self-regulation by delaying the plasma response, and shifting it away from the position of the shards. The ablation might thus be faster, so that more material is deposited earlier along the shard trajectories.

A second area of simplification employed in the present work concerns the geometry and interaction with the structures surrounding the plasma. The geometrical simplifications include the circular plasma cross section, neglect of the toroidicity and the assumption of flux surface homogenised quantities. Relaxing these assumptions would allow for modelling of transient 3D features of the plasma profiles, as well as introducing geometrical order unity corrections to the transport processes involved. For instance, elongating the plasma increases the cross sectional area. This leads to an increase of the time scale for diffusive transport across the plasma cross section, and also decreases the density for a given number of deposited particles. Regarding the surrounding structures, their geometry and conductive properties introduce corrections to the electric field boundary condition. Support for a shaped geometry 
and a finite wall conductivity are implemented in \DREAM, and the sensitivity to these features could therefore be studied in the future.

Finally, another simplification in the present model is the prescribed evolution of the magnetic perturbations. The prescribed magnetic perturbation is sufficient to study qualitative trends involving transport due to magnetic perturbations, as done in this work. However, self-consistent and quantitatively accurate simulations would require coupling to a magnetohydrodynamic (MHD) model, such as JOREK \cite{Nardon_2020,Konsta2021}. The evolution of the magnetic field through the current quench might also be of interest, potentially including deliberately induced perturbations to increase the runaway losses. Even the comparatively smaller magnetic perturbations that may be present during the current quench can have a relatively large impact on the runaway generation and dissipation \cite{Svensson}. This applies especially for cases with an off-axis runaway current profile, as magnetic perturbations induced during the current quench are expected to only partially penetrate into the plasma.  
A major drawback with involving three-dimensional MHD modelling is the significant amount of computational resources required. The thermal quench simulations shown in this work take at most a couple of hours on a desktop computer, even with kinetic electrons, 
while fluid simulations take
a couple of minutes. The orders of magnitude lower computational expense substantially increases the feasibility of exploring a wide range of injection and plasma parameters. 
If the transport processes observed in MHD simulations could be distilled to simplified models suitable for integration in the presented framework, that would allow a major step towards a self-consistent and reliable, as well as  computationally efficient, modelling capability for disruption mitigation schemes.

\section{Conclusions}
\label{conclusions}
An effective disruption mitigation system limits the exposure of the wall to localized transported heat losses and to the impact of high-current runaway electron beams, as well as avoiding excessive forces on the structure. This work evaluates a two-stage deuterium-neon shattered pellet injection scheme, quantifying these three aspects through the maximum runaway current, the thermal quench  timescale, and the transported-to-radiated fraction of thermal energy. 

The study is based on simulations of two-stage SPI in ITER-like plasmas using the \DREAM\ framework, where an initial dilution cooling achieved by deuterium injection is followed by a radiative thermal quench through neon injection. We find that, when the magnetic field breakup is only triggered by the neon pellet, this injection scheme effectively reduces the contribution of transport to the thermal quench, thereby minimizing localised heat loads. This is due to the lower thermal transport rate in the dilutively pre-cooled plasma. 

If the injected neon and deuterium quantities are too low, the current quench time becomes longer than the limit posed by mechanical forces on the wall due to halo currents. The long ohmic current quench time is caused by an incomplete cooling in parts of the plasma, and the associated high conductivity. For large injected densities, however, the simulated current quench times were found to be close to the lower acceptable limit, with a rather weak dependence on the injected densities. 

We found that the two-stage SPI scheme can reduce the hot-tail runaway seed generation by several orders of magnitude. This reduction is explained by the thermalization of the electron distribution between the injections. 

The maximum runaway current exhibits a non-monotonic dependence on the injected deuterium density: At low deuterium densities, an increased deuterium density reduces the avalanche, while at large deuterium densities, the avalanche is enhanced by the higher electric fields and substantial recombination, resulting from the strong radiative cooling. An optimum is found at around an injected deuterium quantity of $1-2\cdot 10^{24}$ atoms, depending on the opacity of the plasma to Lyman radiation. 
Subsequent neon injection in the range of $10^{22}-10^{24}$ atoms gives current quench times and transported fractions of the thermal energy within/close to the respective ranges of acceptable values. 

In  nuclear operation no combination of injected quantities could reduce the maximum runaway current much below $4\,\rm MA$, which is an alarming result. 
The runaway current carried by the beam upon wall impact might be significantly affected in the presence of naturally occurring or externally applied magnetic perturbations remaining after the thermal quench \cite{Konsta2021}, 
which was not considered in this work.

Although the model used in this work comprises an integrated framework accounting for many of the relevant aspects of disruption mitigation by SPI, the various components are treated with different levels of sophistication. For instance, heat and particle transport processes are accounted for in a simplified manner, while runaway electron dynamics is modelled in a comprehensive and state-of-the-art fashion.
The benefits of a two-stage SPI scheme indicated by our results motivate further quantitatively accurate studies.

\section*{Acknowledgements} \noindent The authors are grateful to E.~Nardon, M.~Lehnen and A.~Matsuyama for fruitful discussions.  This work was
supported by the Swedish Research Council (Dnr.~2018-03911). The work
has been carried out within the framework of the EUROfusion
Consortium, funded by the European Union via the Euratom Research and
Training Programme (Grant Agreement No 101052200 — EUROfusion). Views
and opinions expressed are however those of the author(s) only and do
not necessarily reflect those of the European Union or the European
Commission. Neither the European Union nor the European Commission can
be held responsible for them.

\appendix

\section{Overview of {\sc Dream} }
\label{app:dream}

\DREAM\ solves for the flux-surface averaged electron and ion densities, temperatures, as well as the parallel electric field and current. The electron dynamics is resolved fully kinetically or through various fluid models. Going beyond the status of the \DREAM\ framework as represented in reference~\cite{DREAMPaper}, here we overview the aspects of the code relevant for SPI modelling. 

\subsection*{Particle and energy balance}

In the simulations presented here the electron distribution is resolved kinetically for the thermal and superthermal electron populations, while  the highly energetic runaway electrons---above a momentum $p_\mathrm{RE}$ which we here set to $3 m_ec$---are treated as a fluid. Even when the thermal part of the electrons is treated kinetically, fluid energy conservation and quasineutrality equations are also being solved for a Maxwellian bulk electron species to account for radial transport and atomic physics processes. The kinetic collision operator is linearized around this fluid bulk, which enforces that the kinetically resolved thermal electron population remains close to the fluid one, and electron sources on the kinetic grid make sure that the fluid and kinetic particle numbers are consistent.        

The evolution of the energy density $W_\mathrm{M}=3n_\mathrm{M}T_\mathrm{M}/2$ of the Maxwellian bulk electrons is governed by
\begin{eqnarray}
  \frac{\partial W_{\rm M}}{\partial t} &=\sigma_{\rm M}E_{||}^2  +\left(\frac{\partial W_{\rm M}}{\partial t}\right)_\mathrm{ioniz}^\mathrm{abl}+
  \label{eq:heateq}
\frac{1}{r}\frac{\partial}{\partial r}\left[r D_W\frac{\partial T_{\rm M}}{\partial r}\right] \\ \nonumber 
  & - n_{\rm M} \sum_{ij} n_{ij} [L_{ij}(T_{\rm M},\,n_{\rm M})+E_{ij}^\mathrm{ioniz}I_{ij}(T_{\rm M},\,n_{\rm M})]\\ 
&  +
 \int\displaylimits_{p_\mathrm{hot}<p<p_\mathrm{RE}}\Delta\dot{E}_{ee} f \,\mathrm{d}\boldsymbol{p}  - 1.69\cdot 10^{-38} n_{\rm M}^2 \sqrt{T_\mathrm{M}} Z_{\rm eff}. \nonumber
\end{eqnarray}
The conductivity $\sigma _\mathrm{M}$ appearing in the ohmic heating term is calculated using the relativistic expression derived in reference~\cite{Redl_cond}
\begin{equation}
	\sigma _\mathrm{M}=\bar{\sigma}\frac{4\pi\epsilon_0^2T_\mathrm{M}^{3/2}}{Z_\mathrm{eff}\sqrt{m_e e}\ln \Lambda_0},
\label{eq:conductivity_Braams}
\end{equation}
where $\bar{\sigma}(T_\mathrm{M},n_\mathrm{M},Z_\mathrm{eff})$ is calculated by interpolation of the values tabulated in reference~\cite{Redl_cond}.
Here we have also introduced the effective charge $Z_\mathrm{eff}=\sum _{ij} Z_{ij}^2n_{ij}/n_\mathrm{free}$, the dielectric constant $\epsilon _0$, the elementary charge $e$, the thermal Coulomb logarithm $\ln \Lambda _0=14.9-0.5\ln{(n_\mathrm{M}/10^{20})}+\ln{(T_\mathrm{M}/10^3)}$ \cite{wesson}, and the electric field parallel to the magnetic field lines $E_{||}$. 
The second term accounts for the energy required to ionize the ablated pellet material and reach the equilibrium charge state distribution
\begin{eqnarray}
  \left(\frac{\partial W_\mathrm{M}}{\partial t}\right)_\mathrm{ioniz}^\mathrm{abl}&=-\sum _{ij}\Delta E_{ij}^\mathrm{binding}f_{ij}\nonumber \\&\times \sum _{k=1}^{N_\mathrm{s}} \frac{4\pi r_{p,k}^2\dot{r}_{p,k}\rho _{\rm dens}N_\mathrm{A}}{\mathcal{M}}H(r, \rho _{p,k}),
  \label{eq:additionalionizationloss}
\end{eqnarray}
where $\Delta E_{ij}^\mathrm{binding}=\sum _0^{i-1}E_{ij}^\mathrm{ioniz}$ is the total energy required to ionize an atom of species $j$ from neutral to charge state $i$, and the ionization energies $E_{ij}^\mathrm{ioniz}$ are taken from the NIST database\footnote{\url{https://physics.nist.gov/PhysRefData/ASD/ionEnergy.html}}. The assumption of instantaneous equilibration and homogenization over the flux surface, of the ablated material, means that the thermal energy absorbed by the shielding cloud is immediately returned to the background plasma. We therefore do not need any further energy loss terms directly associated with the pellet ablation (assuming the shielding cloud to be optically thick, thus neglecting radiative losses from it).

The third term in equation~\eqref{eq:heateq} describes electron heat diffusion. Here we employ a Rechester-Rosenbluth-type \cite{rechester1978electron} diffusion coefficient $D = \pi q v_{||} R_0 \left(\delta B/B\right)^2$, for particles with a parallel streaming speed of $v_\|$, where the relative amplitude and the parallel correlation length of magnetic perturbations are $\delta B/B$ and $\pi q R_0$, respectively, with $q\approx 1$. Integrating over the Maxwellian electron bulk yields
\begin{eqnarray}
D_W = &\frac{n_{\rm M}}{(\pi^{3/2}v_{T}^3 T_{\rm M})}\\&\times\int \frac{m_e v^2}{2}\left(\frac{v^2}{v_T^2}-\frac{3}{2}\right) D(\boldsymbol{v}) \exp{\left(-\frac{v^2}{v_{T}^2}\right)} \rm d\boldsymbol{v},\nonumber
\label{eq:heatdiff}
\end{eqnarray}
where $v_{T}=\sqrt{2T_{\rm M}/m_e}$ is the electron thermal speed. 
Here, whenever we switch on transport, we prescribe $\delta B/B=7\cdot 10^{-4}$. With $D_W$ evaluated at the initial temperature, this results in a transport loss time scale of a Bessel mode-like decay $a^2/(D_Wx_1^2)$, with $x_1\approx 2.4$, comparable to the expected thermal quench time in ITER \cite{Breizman_review}. 

The fourth term in equation~\eqref{eq:heateq} corresponds to the line radiation and ionization losses. The line radiation rates $L_{ij}(T_\mathrm{M},n_\mathrm{M})$ are taken from the OpenADAS database \cite{ADAS}.
The charge state distribution is calculated by the time dependent rate equations
\begin{eqnarray}
 \left(\frac{\partial n_{ij}}{\partial t}\right)_\mathrm{ioniz}
&= I_{i-1,j}n_{i-1,j}n_{\rm M} -I_{ij} n_{ij}n_{\rm
  M}\nonumber\\& + R_{i+1,j} n_{i+1,j}n_{\rm M} -
R_{ij}n_{ij}n_{\rm M}.
\label{eq:tdre}
\end{eqnarray}
Thus, the total evolution of the ion charge state densities is given by
\begin{equation}
	\frac{\partial n_{ij}}{\partial t}= \left(\frac{\partial n_{ij}}{\partial t}\right)_\mathrm{ioniz}+ \left(\frac{\partial n_{ij}}{\partial t}\right)_\mathrm{SPI}.
\end{equation}
The electron density is determined from the quasi-neutrality condition.

The fifth term in equation \eqref{eq:heateq} describes the collisional energy transfer from hot electrons to the Maxwellian electrons\footnote{As this depends on the energy distribution, which is not resolved for the runaway population; here, only the hot population is accounted for in this interaction term.}. Here $\Delta\dot{E}_{ee} = 4\pi n_\mathrm{M} r_0^2 \ln\Lambda_{ee} m_e c^4/v$, where $r_0 = e^2/(4\pi \varepsilon_0 m_e c^2)$ is the classical electron radius and $\ln \Lambda _{ee}$ is the energy dependent Coulomb logarithm for electron-electron collisions, given in reference~\cite{hesslow2018a}. 
The last term in equation~\eqref{eq:heateq} accounts for the bremsstrahlung losses.

\subsection*{Electric field evolution}

In the cylindrical limit, Faraday's law combined with Amp\`ere's law yield
\begin{equation}
	\mu_0\frac{\partial j_{||}}{\partial t} = \frac{1}{r}\frac{\partial}{\partial r}\left(r\frac{\partial E}{\partial r}\right),
	\label{eq:ampere}
\end{equation}
where $j_{||}$ is the parallel current density, given by the sum of the ohmic current density, 
 and the hot electron and runaway current densities. The ohmic current density is calculated as $j_\mathrm{Ohm}=E_{||}(\sigma_{\rm M} - \sigma _\mathrm{fp}) + \int_{0}^{p_{\rm hot}}ev_{||} f \,\mathrm{d}\boldsymbol{p}$, where $\sigma_{\rm M}$ is the conductivity given in equation~\eqref{eq:conductivity_Braams}. The first term corrects for the part of the conductivity not captured by the conductivity $\sigma _\mathrm{fp}$ resulting from the test-particle Fokker-Planck collision operator used in the kinetic equation. An expression for $\sigma _\mathrm{fp}$ was determined by running numerous DREAM simulations with fixed parameters until the kinetically captured contribution to the ohmic current was equilibrated, and then calculating $\sigma _\mathrm{fp}$ by dividing the ohmic current thus obtained by the fixed value for $E_\mathrm{||}$. This data was used to fit an expression for $\sigma _\mathrm{fp}$ according to
\begin{equation}
	\sigma _\mathrm{fp}=\sigma _\mathrm{M}\left(1+\frac{c_1}{c_2+Z_\mathrm{eff}}\right),
\end{equation}
with $c_1=-1.406$ and $c_2=1.888$. The hot electron current density is $j_\mathrm{hot}=\int_{p_\mathrm{hot}<p<p_\mathrm{RE}}ev_{||} f \,\mathrm{d}\boldsymbol{p}$, and the runaway current density is $j_\mathrm{RE}=ecn_{\rm RE}$. The boundary condition for equation~\eqref{eq:ampere} at $r=a$ is obtained by assuming the plasma to be surrounded by a perfectly conducting wall at $r=b>a$, where the induced electric field is set to $0$. Matching the solution for $r<a$ to the vacuum solution for $a<r<b$ gives $E_{||}(a)=a\ln{(a/b)}\partial E_{||}/\partial r \vert _{r=a}$.

\subsection*{Numerical resolution}
The kinetically resolved electron momentum space, that covers momenta up to the boundary of the numerical runaway region $p_{\rm RE}=3m_e c$, is divided into two regions: Below $p=0.07m_e c$, covering the thermal bulk, the momenta are resolved with $70$ uniform grid cells. Above this, and below $p_{\rm RE}$, momenta are resolved with $50$ uniform grid cells. The pitch angle cosine $\xi$ is resolved by $5$ grid cells uniformly spaced between $-1$ and $1$ in the entire kinetically resolved region. The radial dimension is covered uniformly by $11$ grid cells. The temporal resolution varies between $1-10\,\rm \mu s$, using values from the lower end of this range in the thermal quench phase. The relative tolerance of unknowns, when advancing the system one time step by a nonlinear iterative solver, is set to $10^{-6}$.    

\section{Opacity for deuterium radiation}
\label{app:opacity}
Here we estimate the fraction of the deuterium line radiation which is trapped due to opacity of the plasma in the vicinity of the Lyman lines. We consider a plane, partially ionized, plasma slab, and follow a model described in references~\cite{Morozov} and \cite{Lukash}. The fraction of trapped radiation is determined mainly by two quantities: the optical thickness of the plasma to the line radiation, and the rate of collisional quenching of excited states. If the rate of collisional quenching is low, even an optically thick plasma can be affected by strong radiative energy losses, as the absorption of any photons would quickly be followed by a new photon emission, leading to an efficient radiative transport. 

The optical thickness of a plasma slab of thickness $h$ is given by $\tau=k_{ij}^lh$, with $k_{ij}^l$ being the inverse mean free path of the photon emitted by deexcitation from the $l^\mathrm{th}$ excited state to the ground state of charge state $i$ of ion species $j$ (transitions between excited states are neglected). We will limit ourselves to the first nine excited states due to availability of the necessary data (providing an estimated accuracy of $\sim 1\%$). 
With the natural broadening of the line profile $\gamma$ and the net external broadening $\Gamma$, the inverse photon mean free path is
\begin{equation}
    k_{ij}^l=n_{ij}\frac{(\lambda_{ij}^l)^2}{4\pi}\frac{1}{1+\Gamma/\gamma}.
\end{equation}
Here, $n_{ij}$ is the density of charge state $i$ of ion species $j$ and $\lambda _{ij}^l$ is the wavelength of a photon emitted by deexcitation from the $l^\mathrm{th}$ excited state to the ground state of charge state $i$ of ion species $j$. For a Doppler broadened line, we have
 $   \Gamma/\gamma=1.11\cdot 10^{10}E_{ij}^{0l}\sqrt{m_pT_I/m_I}\nu_{ij}^l$,
where $m_p$ is the proton mass, $m_I$ is the ion (or atom) mass, $T_I$ is the ion temperature, $E_{ij}^{0l}$ is the energy difference between the ground state and the $l^\mathrm{th}$ excited state of charge state $i$ of ion species $j$ and $\nu_{ij}^l$ is the corresponding natural decay rate. Here, all temperatures and energies are given in $\rm eV$, and transition rates in $\rm s^{-1}$.

The effect of collisional quenching of excited states is determined by the ratio of the collisional quenching probability and the decay probability,
\begin{eqnarray}
    \beta _{ij}^l=&\frac{2.7\cdot 10^{-13}n_e}{(E_{ij}^{0l})^3\sqrt{T_e}}\\ \nonumber
    &\times \left[1-\frac{E_{ij}^{0l}}{T_e}\exp{\left(\frac{E_{ij}^{0l}}{T_e}\right)}E_1\left(\frac{E_{ij}^{0l}}{T_e}\right)\right],
\end{eqnarray}
where $E_1(x)=\int_x^\infty \mathrm{d}t \exp{(-t)}/t$ is the integral exponent. The fraction of radiation escaping from a plane ion slab for a line emitted by deexcitation of the $l^\mathrm{th}$ excited state from charge state $i$ of ion species $j$ is then given by
 $   B_{ij}^l=W_{ij}^l/(\beta_{ij}^l+1)$,
where 
 $   W_{ij}^l=(1+\beta_{ij}^l)P_\mathrm{a}/(\beta_{ij}^l+P_\mathrm{a})$,
and
   $ P_\mathrm{a}=[1+\tau\sqrt{\pi\ln{(\tau+1)}}]^{-1}$
is the probability for a photon to travel a distance $h$ without being absorbed (the ``1'' in the denominator is included \textit{ad-hoc} in the expression valid in the optically thick limit, to make sure that $P_\mathrm{a}\rightarrow 1$ as it should for thin plasma slabs). 

To find the fraction of the total radiation that escapes the plasma, we also need to know the relative intensity of the different lines. The intensity distribution over the different lines is very different for excited states populated by excitations from lower states (dominantly the ground state) compared to excited states populated by recombination. The reason for this is that excitations from lower states primarily populate the lower excited states, while recombination populates higher excited states. In addition, part of the potential energy change is radiated away during a free-bound transition, resulting in a continuous spectrum to which the plasma is essentially transparent.

For radiation following excitation from lower states, we calculate the relative line intensities based on data from reference~\cite{transition_data}, according to
 $   L_{ij}^l\propto E_{ij}^{0l}\phi _{ij}^l$,
where 
 $   \phi _{ij}^l=r_{ij}^l\bar{n}_\mathrm{Saha}(l)\nu_{ij}^l=r_{ij}^l(l+1)^2\exp{(E_{ij}^{l\infty}/T_e)}\nu_{ij}^l$
is proportional to the relative occupation $\bar{n}_\mathrm{Saha}$ of the $l^\mathrm{th}$ excited state at Saha equilibrium, and $E_{ij}^{l\infty}$ is the ionization energy from the $l^\mathrm{th}$ excited state. The coefficients $r_{ij}^l$ (tabulated in~\cite{transition_data}) describe the deviation from the Saha equilibrium prevailing in a dense enough plasma to reach local thermodynamic equilibrium. The transition rates $\nu_{ij}^l$ are tabulated in~\cite{Verner_atomic}. The fraction of line radiation following excitations from the ground state escaping the plasma can now be expressed as
\begin{equation}
    f_\mathrm{esc}=\frac{\sum _{ij} n_{ij}\sum _l B_{ij}^lL_{ij}^l}{\sum _{ij} n_{ij}\sum _l L_{ij}^l}=\frac{\sum _l B_{0D}^lE_{0D}^{0l}\phi _{0D}^l}{\sum _l E_{0D}^{0l}\phi _{0D}^l},
	\label{eq:f_esc}
\end{equation}
where the second equality holds for hydrogen isotopes where there is only one radiating charge state. 

When the excited states are populated by recombination, energy will also be radiated away during the free-bound transition, in addition to the subsequent deexcitation through bound-bound transitions. We consider two types of recombination event: one with transitions directly from a free state to the bound ground state, and one consisting of a free-bound transition to an excited bound state, followed by a direct transition to the ground state. If the recombination is radiative, a transition from state $l$ (which can be the ground state, $l=0$) will be preceded by a transition from the continuum to state $l$, while releasing an average photon of energy $E_{ij,\mathrm{rec}}^l\approx E_{ij}^{\infty 0}+T_e-E_{ij}^{0l}$. The free-bound transitions give rise to a continuous spectrum, to which the plasma is essentially transparent. The resulting escape factor for recombination radiation becomes
\begin{eqnarray}
    f_\mathrm{esc}&=\frac{\sum _{ij} n_{ij}\sum _l (B_{ij}^lL_{ij}^l+L_{ij,\mathrm{rec}}^l)}{\sum _{ij} n_{ij}\sum _l (L_{ij}^l+L_{ij,\mathrm{rec}}^l)}\nonumber\\&=\frac{\sum _l (B_{0D}^lE_{0D}^{0l}+E_{0D,\mathrm{rec}}^l)\phi_{0D}^l}{\sum _l (E_{0D}^{0\infty}+T_e)\phi _{0D}^l}.
    \label{eq:f_esc_rec}
\end{eqnarray}
As the coefficients in~\cite{transition_data} only apply to recombination events involving an excited bound state, we replace $\phi _0^l$ in equation~\eqref{eq:f_esc_rec} with the transition rates from~\cite{Seaton},  which also include radiative recombination directly to the ground state.

\begin{figure}
	\centering

		\includegraphics[width=\columnwidth]{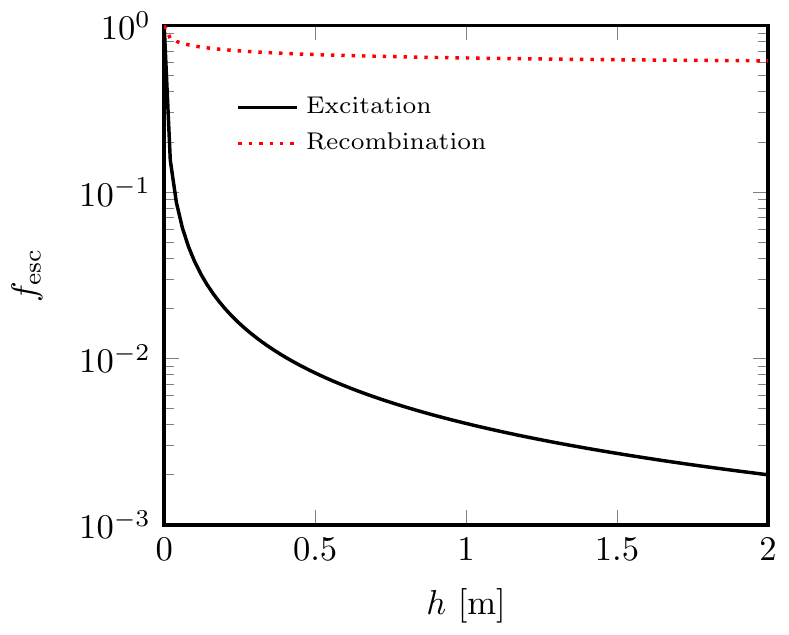}
	\caption{Escaping fraction of deuterium radiation as a function of slab thickness, for a plasma with electron density $n_e=10^{20}\,\rm  m^{-3}$, neutral deuterium density $n_D=4\cdot 10^{21}\,\rm m^{-3}$, and temperature $T_e=1.38\,\rm eV$. Black solid: Radiation following excitation from the ground state. Dotted red: recombination radiation.}
	\label{fig:f_esc}
\end{figure}

The escape factor as a function of the slab thickness $h$ is shown in figure~\ref{fig:f_esc}, for $h$ up to the minor radius  of an ITER-like plasma $a=2\,\rm   m$. Results are shown for both radiation following excitations from the ground state (solid), from equation~\eqref{eq:f_esc}, and for recombination radiation (dotted), from equation~\eqref{eq:f_esc_rec}. The plasma parameters considered, $n_e=10^{20}\,\rm m^{-3}$ and $T_e=1.38\,\rm eV$, and neutral deuterium density $n_{D}=4\cdot 10^{21}\,\rm m^{-3}$, are chosen to match the ones for which there are tabulated data in reference~\cite{transition_data}. One can see that the line radiation following excitations is below $1\%$ already for $h=0.5\,\rm m$. For the recombination radiation, on the other hand, the escaping fraction remains above $60\%$ for the $h$-range shown.

\bibliography{main.bib}

\providecommand{\newblock}{}
\begin{thebibliography}{10}
\expandafter\ifx\csname url\endcsname\relax
  \def\url#1{{\tt #1}}\fi
\expandafter\ifx\csname urlprefix\endcsname\relax\def\urlprefix{URL }\fi
\providecommand{\eprint}[2][]{\url{#2}}

\bibitem{RP97}
Rosenbluth M and Putvinski S 1997 {\em Nuclear Fusion\/} {\bf 37} 1355
  \urlprefix\url{https://doi.org/10.1088/0029-5515/37/10/i03}

\bibitem{ITERDisruptions}
Hollmann E, Aleynikov P, F{\"u}l{\"o}p T, Humphreys D, Izzo V, Lehnen M, Lukash
  V, Papp G, Pautasso G, Saint-Laurent F {\em et~al.\/} 2015 {\em Physics of
  Plasmas\/} {\bf 22} {021802--1}
  \urlprefix\url{https://doi.org/10.1063/1.4901251}

\bibitem{LehnenIAEA}
Lehnen M, Jachmich S, Kruezi U and the ITER DMS~task force 2020 The {ITER}
  disruption mitigation strategy
  \url{https://conferences.iaea.org/event/217/contributions/17867/} presented
  by M. Lehnen at the IAEA Technical Meeting on Plasma Disruption and their
  Mitigation

\bibitem{Chiu_1998}
Chiu S, Rosenbluth M, Harvey R and Chan V 1998 {\em Nuclear Fusion\/} {\bf 38}
  1711 \urlprefix\url{https://doi.org/10.1088%2F0029-5515%2F38%2F11%2F309}

\bibitem{Harvey2000}
Harvey R~W, Chan V~S, Chiu S~C, Evans T~E, Rosenbluth M~N and Whyte D~G 2000
  {\em Physics of Plasmas\/} {\bf 7} 4590
  \urlprefix\url{https://doi.org/10.1063/1.1312816}

\bibitem{Nardon_2020}
Nardon E, Hu D, Hoelzl M and and D~B 2020 {\em Nuclear Fusion\/} {\bf 60}
  126040 \urlprefix\url{https://doi.org/10.1088/1741-4326/abb749}

\bibitem{MS2017}
Martin-Sol\'{i}s J~R, Loarte A and Lehnen M 2017 {\em Nuclear Fusion\/} {\bf
  57} 066025 \urlprefix\url{https://doi.org/10.1088/1741-4326/aa6939}

\bibitem{ITER_OV}
Vallhagen O, Embreus O, Pusztai I, Hesslow L and Fülöp T 2020 {\em Journal of
  Plasma Physics\/} {\bf 86} 475860401
  \urlprefix\url{https://doi.org/10.1017/S0022377820000859}

\bibitem{AB17}
Aleynikov P and Breizman B~N 2017 {\em Nuclear Fusion\/} {\bf 57} 046009
  \urlprefix\url{https://doi.org/10.1088/1741-4326/aa5895}

\bibitem{Svenningsson2021}
Svenningsson I, Embreus O, Hoppe M, Newton S~L and F\"ul\"op T 2021 {\em
  Physical Review Letters\/} {\bf 127}(3) 035001
  \urlprefix\url{https://link.aps.org/doi/10.1103/PhysRevLett.127.035001}

\bibitem{DREAMPaper}
Hoppe M, Embreus O and Fülöp T 2021 {\em Computer Physics Communications\/}
  {\bf 268} 108098 \urlprefix\url{https://doi.org/10.1016/j.cpc.2021.108098}

\bibitem{Parks_TSDW}
Parks P 2017 A theoretical model for the penetration of a shattered-pellet
  debris plume
  \url{https://tsdw.pppl.gov/Talks/2017/Lexar/Wednesday\%20Session\%201/Parks.pdf}
  presented at the Theory and Simulation of Disruptions Workshop

\bibitem{Fulop20}
Fülöp T, Helander P, Vallhagen O, Embréus O, Hesslow L, Svensson P, Creely
  A~J, Howard N~T and Rodriguez-Fernandez P 2020 {\em Journal of Plasma
  Physics\/} {\bf 86} 474860101
  \urlprefix\url{https://doi.org/10.1017/S002237782000001X}

\bibitem{HesslowNF}
Hesslow L, Embréus O, Vallhagen O and Fülöp T 2019 {\em Nuclear Fusion\/}
  {\bf 59} 084004 \urlprefix\url{https://doi.org/10.1088/1741-4326/ab26c2}

\bibitem{ParksGA2016}
Parks P 2016 Modeling dynamic fracture of cryogenic pellets Tech. Rep.
  GA-A28352 General Atomics \urlprefix\url{https://www.osti.gov/biblio/1344852}

\bibitem{DiHu}
Hu D, Nardon E, Lehnen M, Huijsmans G and van Vugt~and D 2018 {\em Nuclear
  Fusion\/} {\bf 58} 126025
  \urlprefix\url{https://doi.org/10.1088/1741-4326/aae614}

\bibitem{AkinobuREM}
Matsuyama A, Nardon E, Honda M and Lehnen M 2017 {ITER SPI} modelling for
  runaway electron avoidance
  \url{https://conferences.iaea.org/event/217/contributions/16702/} presented
  at the Theory and Simulation of Disruptions Workshop

\bibitem{NardonIAEA}
Nardon E, Matsuyama A and Lehnen M 2020 On the possible injection schemes with
  the {ITER SPI} system
  \url{https://conferences.iaea.org/event/217/contributions/16702/} presented
  at the IAEA Technical Meeting on Plasma Disruption and their Mitigation

\bibitem{Parks_Thurnbull}
Parks P~B and Turnbull R~J 1978 {\em The Physics of Fluids\/} {\bf 21} 1735
  (\textit{Preprint}
  \eprint{https://aip.scitation.org/doi/pdf/10.1063/1.862088})
  \urlprefix\url{https://aip.scitation.org/doi/abs/10.1063/1.862088}

\bibitem{Kiramov_2020}
Kiramov D~I and Breizman B~N 2020 {\em Nuclear Fusion\/} {\bf 60} 084004
  \urlprefix\url{https://doi.org/10.1088/1741-4326/ab966a}

\bibitem{stangeby2000plasma}
Stangeby P 2000 {\em The Plasma Boundary of Magnetic Fusion Devices\/} Series
  in Plasma Physics and Fluid Dynamics (Taylor \& Francis) ISBN 9780750305594
  \urlprefix\url{https://books.google.se/books?id=qOliQgAACAAJ}

\bibitem{ShirakiIAEA}
Shiraki D, Herfindal J, Baylor L~R, Hollmann E~M, Lasnier C, Bykov I, Eidietis
  N, Raman R, Sweeney R, Sheikh U, Gerasimov S, Jachmich S, Lehnen M, Kim J,
  Jang J~J, Meitner S and Gebhart T 2020 Particle assimilation during shattered
  pellet injection
  \url{https://conferences.iaea.org/event/217/contributions/16713/} presented
  at the IAEA Technical Meeting on Plasma Disruption and their Mitigation

\bibitem{Lengyel_1999}
Lengyel L, Büchl K, Pautasso G, Ledl L, Ushakov A, Kalvin S and Veres G 1999
  {\em Nuclear Fusion\/} {\bf 39} 791
  \urlprefix\url{https://doi.org/10.1088/0029-5515/39/6/307}

\bibitem{ADAS}
Summers H~P 2004 The {ADAS} user manual, version 2.6
  \url{http://www.adas.ac.uk}

\bibitem{Pshenov_2019}
Pshenov A, Kukushkin A, Marenkov E and Krasheninnikov S 2019 {\em Nuclear
  Fusion\/} {\bf 59} 106025
  \urlprefix\url{https://doi.org/10.1088/1741-4326/ab3144}

\bibitem{Morozov}
Morozov D, Baronova E and Senichenkov I 2007 {\em Plasma Physics Reports\/}
  {\bf 33} 906 \urlprefix\url{https://doi.org/10.1134/S1063780X07110037}

\bibitem{Lukash}
Lukash V, Mineev A and Morozov D 2007 {\em Nuclear Fusion\/} {\bf 47} 1476
  \urlprefix\url{https://doi.org/10.1088/0029-5515/47/11/009}

\bibitem{Pegourie_2004}
P{\'{e}}gouri{\'{e}} B, Waller V, Dumont R~J, Eriksson L~G, Garzotti L,
  G{\'{e}}raud A and Imbeaux F 2004 {\em Plasma Physics and Controlled
  Fusion\/} {\bf 47} 17--35
  \urlprefix\url{https://doi.org/10.1088/0741-3335/47/1/002}

\bibitem{Gilliard_1980}
Gilliard R~P and Kim K 1980 {\em IEEE Transactions on Plasma Science\/} {\bf 8}
  477--484

\bibitem{Pegourie_2007}
P{\'{e}}gouri{\'{e}} B 2007 {\em Plasma Physics and Controlled Fusion\/} {\bf
  49} 87 \urlprefix\url{https://doi.org/10.1088%2F0741-3335%2F49%2F8%2Fr01}

\bibitem{Rozhansky_2004}
Rozhansky V, Senichenkov I, Veselova I and Schneider R 2004 {\em Plasma Physics
  and Controlled Fusion\/} {\bf 46} 575--591
  \urlprefix\url{https://doi.org/10.1088/0741-3335/46/4/001}

\bibitem{Konsta2021}
S\"{a}rkim\"{a}ki K, Artola J and Hoelzl M 2021 Confinement of passing and
  trapped runaway electrons in the simulation of an {ITER} current quench,
  \emph{to be submitted to Nuclear~Fusion}

\bibitem{Svensson}
Svensson P, Embreus O, Newton S~L, Särkimäki K, Vallhagen O and Fülöp T
  2021 {\em Journal of Plasma Physics\/} {\bf 87} 905870207
  \urlprefix\url{https://doi.org/10.1017/S0022377820001592}

\bibitem{Redl_cond}
Redl A, Angioni C, Belli E and Sauter O 2021 {\em Physics of Plasmas\/} {\bf
  28} 022502 (\textit{Preprint} \eprint{https://doi.org/10.1063/5.0012664})
  \urlprefix\url{https://doi.org/10.1063/5.0012664}

\bibitem{wesson}
Wesson J 2011 {\em Tokamaks\/} 4th ed (Oxford, UK: Oxford university press)

\bibitem{rechester1978electron}
Rechester A~B and Rosenbluth M~N 1978 {\em Phys. Rev. Lett.\/} {\bf 40}(1) 38
  \urlprefix\url{https://link.aps.org/doi/10.1103/PhysRevLett.40.38}

\bibitem{Breizman_review}
Breizman B~N, Aleynikov P, Hollmann E~M and Lehnen M 2019 {\em Nuclear
  Fusion\/} {\bf 59} 083001
  \urlprefix\url{https://doi.org/10.1088/1741-4326/ab1822}

\bibitem{hesslow2018a}
Hesslow L, Embr{\'e}us O, Hoppe M, DuBois T, Papp G, Rahm M and F{\"u}l{\"o}p T
  2018 {\em Journal of Plasma Physics\/} {\bf 84} 905840605
  \urlprefix\url{https://doi.org/10.1017/S0022377818001113}

\bibitem{transition_data}
Johnson L and Hinnov E 1973 {\em Journal of Quantitative Spectroscopy and
  Radiative Transfer\/} {\bf 13} 333 ISSN 0022-4073
  \urlprefix\url{http://www.sciencedirect.com/science/article/pii/0022407373900642}

\bibitem{Verner_atomic}
Verner D, Verner E and Ferland G 1996 {\em Atomic Data and Nuclear Data
  Tables\/} {\bf 64} 1 ISSN 0092-640X
  \urlprefix\url{http://www.sciencedirect.com/science/article/pii/S0092640X96900182}

\bibitem{Seaton}
Seaton M~J 1959 {\em Monthly Notices of the Royal Astronomical Society\/} {\bf
  119} 81 ISSN 0035-8711 (\textit{Preprint}
  \eprint{https://academic.oup.com/mnras/article-pdf/119/2/81/8078011/mnras119-0081.pdf})
  \urlprefix\url{https://doi.org/10.1093/mnras/119.2.81}

\end{thebibliography}

\end{document}